\documentclass[journal]{IEEEtran}
\usepackage{amsmath}
\usepackage{algorithmic}
\usepackage[T1]{fontenc}
\usepackage{algorithm}
\usepackage{array}
\usepackage{bm}
\usepackage{color}
\usepackage{flushend}
\ifCLASSOPTIONcompsoc
\usepackage[caption=false, font=normalsize, labelfont=sf, textfont=sf]{subfig}
\else
\usepackage[caption=false, font=footnotesize]{subfig}
\usepackage{textcomp}
\usepackage{booktabs}
\usepackage{multicol}
\usepackage{stfloats}
\usepackage{url}
\usepackage{verbatim}
\usepackage{graphicx}
\usepackage{booktabs}
\usepackage{cite}
\usepackage[table,xcdraw]{xcolor}
\usepackage{booktabs}
\usepackage{multirow}

\usepackage{hyperref}
\hypersetup{
     colorlinks = true,
     linkcolor = blue,
     anchorcolor = red,
     citecolor = magenta,
     filecolor = blue,
     urlcolor = black,
}

\usepackage{xcolor}
\definecolor{babypink}{rgb}{0.96, 0.76, 0.76}
\definecolor{flamingopink}{rgb}{0.99, 0.56, 0.67}
\definecolor{bananamania}{rgb}{0.98, 0.91, 0.71}
\definecolor{cambridgeblue}{rgb}{0.64, 0.76, 0.68}
\definecolor{asparagus}{rgb}{0.53, 0.66, 0.42}
\definecolor{desertsand}{rgb}{0.93, 0.79, 0.69}
\definecolor{tropicalholiday}{HTML}{8ECFC9}
\definecolor{ao(english)}{rgb}{0.0, 0.5, 0.0}

\newenvironment{packeditemize}{
	\begin{list}{$\bullet$}{
			\setlength{\labelwidth}{4pt}
			\setlength{\itemsep}{0pt}
			\setlength{\leftmargin}{\labelwidth}
			\addtolength{\leftmargin}{\labelsep}
			\setlength{\parindent}{0pt}
			\setlength{\listparindent}{\parindent}
			\setlength{\parsep}{0pt}
			\setlength{\topsep}{1pt}}}{\end{list}}

\begin{document}

\title{Fast High-Quality Enhanced Imaging Algorithm for Layered Dielectric Targets Based on MMW MIMO-SAR System}

\author{Xu Chen,
        Guangsheng Yu\textbf{},
        Zhian Yuan\textbf{},
        Hao Wu\textbf{},
        Yilin Jiang\textbf{},
        Ying Wang\textbf{},
        Bin Deng\textbf{},
       and Limin Guo
\thanks{Manuscript received November xx, 2024; This work was supported by xxx. \textit{(Corresponding author: Limin Guo.)}}
\thanks{X. Chen, Y. Jiang and L. Guo are with the College of Information and Communication Engineering, Harbin Engineering University, Harbin 150001, China. (e-mail: xuchen95909@126.com; jiangyilin@hrbeu.edu.cn; guolimin@hrbeu.edu.cn;).}
\thanks{G. Yu is with the University of Technology Sydney, Australia. (e-mail: guangsheng.yu@uts.edu.au;).}
\thanks{Z. Yuan and B. Deng are with the National University of Defense Technology, Changsha 410073, China. (e-mail: yuanzhian@nudt.edu.cn; dengbin@nudt.edu.cn;).}
\thanks{H. Wu is with the Institute of Dataspace, Hefei 230000, China. (e-mail: jsczjtwh@126.com;).}
\thanks{Y. Wang is with the Hangzhou Dianzi University, Hangzhou 310018, China. (e-mail: 90023@hdu.edu.cn;).}
}

\markboth{}%
{Shell \MakeLowercase{\textit{et al.}}: A Sample Article Using IEEEtran.cls for IEEE Journals}

\maketitle
\begin{abstract}
Millimeter-wave (MMW) multiple-input multiple-output synthetic aperture radar (MIMO-SAR) system is a technology that can achieve high resolution, high frame rate, and all-weather imaging and has received extensive attention in the non-destructive testing and internal imaging applications of layered dielectric targets. However, the non-ideal scattering effect caused by dielectric materials can significantly deteriorate the imaging quality when using the existing MIMO-SAR fast algorithms. This paper proposes a rapid, high-quality dielectric target-enhanced imaging algorithm for a new universal non-uniform MIMO-SAR system. The algorithm builds on the existing non-uniform MIMO-SAR dielectric target frequency-domain algorithm (DT-FDA) by constructing a forward sensing operator and incorporating it into the alternating direction method of multipliers (ADMM) framework. This approach avoids large matrix operations while maintaining computational efficiency. By integrating an optimal regularization parameter search, the algorithm enhances the image reconstruction quality of dielectric internal structures or defects. Experimental results show the proposed algorithm outperforms IBP and DT-FDA, achieving better focusing, sidelobe suppression, and 3D imaging accuracy. It yields the lowest image entropy (8.864) and significantly improves efficiency (imaging time: 15.29 s vs. 23295.3 s for IBP).

\end{abstract}


\section{Introduction}\label{sec:intro}

\IEEEPARstart{I}{N} recent years, non-metallic dielectric materials like polytetrafluoroethylene (PTFE) have gained widespread use in various military and civilian applications due to their exceptional physico-chemical properties~\cite{1,2}. These materials are often manufactured as components with layered structures, such as aerospace insulation layers and bridge seismic pads. However, structural flaws resulting from imperfect manufacturing processes or material aging can significantly degrade component performance, potentially leading to safety hazards. Consequently, there is a critical need for non-destructive testing and internal imaging of layered dielectric targets to evaluate their performance and ensure safety~\cite{3,4}.

Millimeter-wave (MMW) multiple-input multiple-output synthetic aperture radar (MIMO-SAR) imaging technology addresses these challenges using wideband MMW signals and a real-aperture linear MIMO array with mechanical scanning. This enables three-dimensional (3D) reconstruction of spatial targets~\cite{5,6,7,8}, offering simplified implementation, reduced hardware costs, high-resolution imaging, robust penetration of opaque materials, and enhanced multi-angle information capture~\cite{9,10,11,12}.
Optimized methods, such as the improved back-projection (IBP) algorithm~\cite{19} and frequency-domain processing techniques~\cite{20}, have significantly advanced dielectric target imaging. These methods address the limitations of traditional MIMO-SAR fast imaging algorithms~\cite{14,15,16,17,18}, which struggle with changes in electromagnetic wave propagation and spatial wave number due to spatial discontinuities. By utilizing spatial-spectral resources and multi-angle information, these approaches achieve high imaging accuracy and are compatible with various MIMO antenna array configurations.

\smallskip
\noindent\textbf{Research gaps. }The development of imaging algorithms based on free space is approaching maturity~\cite{36,37,38,39,40}. However, existing methods in MMW MIMO-SAR imaging for layered dielectric targets still face significant challenges, despite advancing the field. 
\begin{packeditemize}
    \item \textbf{Computational Efficiency.} Conventional temporal domain algorithms such as IBP~\cite{19}, despite their accuracy and versatility, struggle with the computational demands of high-resolution MMW imaging. The increased mesh points and high-dimensional echo data from MIMO-SAR systems severely limit their efficiency. 

    \item \textbf{Image Quality.} Fast imaging algorithms have emerged to address this issue, but they often rely on uniform MIMO array topologies~\cite{20}, overlooking the potential benefits of optimized non-uniform arrays in suppressing grating sidelobes~\cite{21,22,23,24}. Furthermore, current imaging techniques, including the state-of-the-art (SOTA) frequency-domain approaches, still grapple with image quality issues at high dynamic ranges. These include noticeable clutter and sidelobe effects and non-ideal scattering from dielectric materials, which impede subsequent target recognition and image interpretation~\cite{25}. 

    \item \textbf{Resource Requirements.} While compressive sensing (CS) methods show promise for enhancing image quality~\cite{26,27}, they typically require handling massive measurement matrices during iterative solving, placing extreme demands on computing resources. Addressing these gaps is crucial for advancing the field of dielectric target imaging using non-uniform MIMO-SAR systems.
\end{packeditemize}

\smallskip
\noindent\textbf{Contributions. }In this paper, we propose a fast and high-quality dielectric target-enhanced imaging algorithm based on the non-uniform MIMO-SAR system. Our approach integrates the existing non-uniform MIMO-SAR dielectric target frequency-domain imaging algorithm (DT-FDA)~\cite{25} into an alternating direction method of multipliers (ADMM)-based iterative solution process, which avoids the large-scale matrix storage and inversion processes in traditional ADMM iterative solution framework, reducing the computing resource requirements while ensuring similar high computational efficiency as DT-FDA. In addition, the good sparse performance of the algorithm achieves an effective improvement in image quality.

The main contributions of this paper are as follows:

\begin{packeditemize}
    \item We establish a forward sensing operator based on the existing DT-FDA algorithm, replacing the large-scale sensing matrix in the echo signal model. This approach significantly reduces computational complexity and memory requirements, making it more suitable for processing non-uniform MIMO-SAR data.

    \item We optimize the iterative solution model of the ADMM to avoid the storage and inversion of large-scale sensing matrices in existing CS methods. By incorporating our established forward sensing operator into the ADMM framework, we eliminate the need for huge matrix storage and inversion operations, thus improving computational efficiency.

    \item We combine the algorithm with an existing fast optimal regularization parameter search approach to achieve high-quality enhanced imaging of the dielectric target's internal structure while ensuring high computational efficiency. This integration allows for better reconstruction of internal structures or defects in dielectric targets, balancing image quality with processing speed.
\end{packeditemize}

Experimental results validate the proposed algorithm's superior performance in focusing, sidelobe suppression, and non-ideal scattering mitigation. It achieves the lowest Image Entropy (IE) of 8.864, outperforming IBP and DT-FDA in 3D imaging. The algorithm's efficiency is notable, reducing imaging time from 23295.3 s--IBP to 15.29 s, comparable to DT-FDA's 14.51 s. Comparative analysis confirms consistently sharper images and lower IE values, indicating enhanced focus and artifact reduction. These improvements, with potential GPU acceleration, position the algorithm for real-time enhanced imaging applications.

\smallskip
\noindent\textbf{Paper organization. }The remainder of this paper is organized as follows. Section~\ref{sec:background} presents the system model and problem formulation. Section~\ref{sec:design} details the proposed fast and high-quality dielectric target-enhanced imaging algorithm. Experimental results and performance analysis are provided in Section~\ref{sec:evaluation}, and the related work is in Section~\ref{sec:related}. Section~\ref{sec:conclusion} concludes the paper.

\section{System Model: Non-Uniform MIMO-SAR DT-FDA}\label{sec:background}

\begin{figure}[t!]
	\centering
	\includegraphics[width=0.9\linewidth]{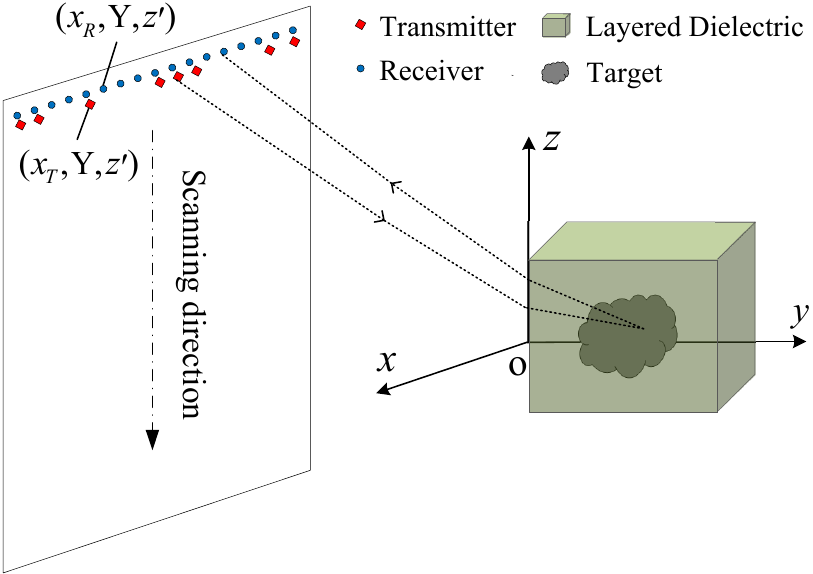}
	\caption{Imaging configuration of layered dielectric targets using non-uniform MIMO-SAR system
	}
	\label{fig: 5.2}
\end{figure}

Fig. \ref{fig: 5.2} presents the imaging configuration of the non-uniform MIMO-SAR system for layered dielectric targets. Under the spatial Cartesian coordinate system $\text{o-}xyz$, the MIMO-SAR aperture lies on the $y=\text{Y}$ plane. The one-dimensional MIMO array is positioned along the $x$ direction and mechanically traverses along the $z$ direction, thereby equivalently creating a two-dimensional (2D) aperture to illuminate the dielectric target area. The imaging target is the internal structure or defect in the uniform layered dielectric material. The relative permittivity of the dielectric material is $\varepsilon$, and the air-dielectric interface is positioned on the $y=0$ plane.

Assuming that the polarization directions of the radar transmitter and receiver are the same, and not considering the attenuation of signal propagation, the target echo signal expression can be derived based on the first-order Born approximation theory\cite{31} as follows:
\begin{equation}
	\begin{aligned}
		y(k,{\bm{b}_{T}},{\bm{b}_{R}})=\text{j}{{\eta }_{0}}k\int_{\Omega }{\sigma (\bm{b})\cdot {{A}_{TR}}(k,\bm{b},{\bm{b}_{T}},{\bm{b}_{R}})\text{d}\bm{b}},
	\end{aligned}
	\label{eq: 5.1}
\end{equation}
where ${\bm{b}_{T}}=({{x}_{T}},\text{Y},{z}')$, ${\bm{b}_{R}}=({{x}_{R}},\text{Y},{z}')$ and $\bm{b}=(x,y,z)$ separately denote the coordinates of    the transmitter, the receiver and  the imaging target. The wave impedance in free space is represented by ${\eta }_{0}$. The wave number of the electromagnetic wave in free space is $k={2\text{ }\!\!\pi\!\!\text{ }f}/{\text{c}}$, where $f$ and $\text{c}$ are the carrier-frequency and the velocity of light, respectively. $\sigma (\bm{b})$ and $\Omega $ represent the target's reflection coefficient and the image reconstruction region, respectively. ${{A}_{TR}}(k,\bm{b},{\bm{b}_{T}},{\bm{b}_{R}})$ is the two-way dyadic Green's function\cite{32}, that is: 
\begin{equation}
	\begin{aligned}
		{{A}_{TR}}(k,\bm{b},{\bm{b}_{T}},{\bm{b}_{R}})={{A}_{T}}(k,\bm{b},{\bm{b}_{T}})\cdot {{A}_{R}}(k,\bm{b},{\bm{b}_{R}})
	\end{aligned},
	\label{eq: 5.2}
\end{equation}
where ${{A}_{T}}(k,\bm{b},{\bm{b}_{T}})$ and ${{A}_{R}}(k,\bm{b},{\bm{b}_{R}})$ respectively correspond to the one-way dyadic Green's functions from transmitter and receiver to the target scattering point:
\begin{equation}
	\begin{aligned}
		{{A}_{T}}(k,\bm{b},{\bm{b}_{T}})=\exp (-\text{j}(k{{R}_{TA}}+{{k}_{\text{ }\!\!\varepsilon\!\!\text{ }}}{{R}_{TM}})),
	\end{aligned}
	\label{eq: 5.3}
\end{equation}
\begin{equation}
	\begin{aligned}
		{{A}_{R}}(k,\bm{b},{\bm{b}_{R}})=\exp (-\text{j}(k{{R}_{RA}}+{{k}_{\text{ }\!\!\varepsilon\!\!\text{ }}}{{R}_{RM}})),
	\end{aligned}
	\label{eq: 5.4}
\end{equation}
where ${{k}_{\text{ }\!\!\varepsilon\!\!\text{ }}}=\sqrt{\text{ }\!\!\varepsilon\!\!\text{ }}k$ is the wave number of electromagnetic wave in the dielectric medium. ${{R}_{TA}}$ and ${{R}_{TM}}$ are the distances of the transmitted electromagnetic wave propagating in air and the medium, respectively. ${{R}_{RA}}$ and ${{R}_{RM}}$ are the distances of the reflected electromagnetic echoes of the target propagating in air and the medium, respectively. Let $(x_{bT},0,z_{bT})$ and $(x_{bR},0,z_{bR})$ represent the refraction point coordinates of the transmitted and received electromagnetic waves on the air-medium interface, respectively, and we can obtain:
\begin{equation}
	\left\{ \begin{aligned}
		&
		R_{TA}=\sqrt{\left(x_T-x_{bT}\right)^2+\mathrm{Y}^2+\left(z^{\prime}-z_{bT}\right)^2},\\
		&
		R_{TM}=\sqrt{\left(x_{bT}-x\right)^2+y^2+\left(z_{bT}-z\right)^2},\\
		&
		R_{RA}=\sqrt{\left(x_R-x_{bR}\right)^2+\mathrm{Y}^2+\left(z^{\prime}-z_{bR}\right)^2},\\
		&
		R_{RM}=\sqrt{\left(x_{bR}-x\right)^2+y^2+\left(z_{bR}-z\right)^2}.
	\end{aligned} \right.
	\label{eq: 5.5}
\end{equation}

According to the electromagnetic theory, the spatial wave number component of electromagnetic waves parallel to the air-medium boundary (i.e., the MIMO-SAR array plane) is continuous in different media. By applying the stationary phase principle to decompose the exponential terms in~\eqref{eq: 5.3} and \eqref{eq: 5.4} into plane waves, we can arrive at:
\begin{equation}
	\begin{aligned}
		{{A}_{T}}(k,\bm{b},{\bm{b}_{T}})& \approx \int{\exp (-\text{j}{{k}_{xYT}}\sqrt{{{({{x}_{T}}-{{x}_{bT}})}^{2}}+{{\text{Y}}^{2}}})} \\ 
		& \text{ }\text{ }\times \exp (-\text{j}{{k}_{xyT}}\sqrt{{{({{x}_{bT}}-x)}^{2}}+{{y}^{2}}}) \\ 
		& \text{ }\text{ }\times \exp (\text{j}{{k}_{z}}({z}'-z)/2)\text{d}{{k}_{z}}, 
	\end{aligned}
	\label{eq: 5.6}
\end{equation}
\begin{equation}
	\begin{aligned}
		{{A}_{R}}&(k, \bm{b},{\bm{b}_{R}})\approx \iint{\exp (\text{j}({{k}_{yR0}}\text{Y}-{{k}_{yR1}}y))} \\ 
		& \text{ }\text{ }\text{ }\text{ }\text{ }\text{ }\text{ }\text{ }\text{ }\text{ }\text{ }\text{ }\text{ }\text{ }\text{ }\text{ }\text{ }\times \exp (\text{j}{{k}_{xR}}({{x}_{R}}-x))\\
		& \text{ }\text{ }\text{ }\text{ }\text{ }\text{ }\text{ }\text{ }\text{ }\text{ }\text{ }\text{ }\text{ }\text{ }\text{ }\text{ }\text{ } \times \exp (\text{j}{{k}_{z}}({z}'-z)/2)\text{d}{{k}_{xR}}\text{d}{{k}_{z}},
	\end{aligned}
	\label{eq: 5.7}
\end{equation}
where ${{x}_{bT}}$ represents the abscissa of the refraction point of the transmitted signal propagating to the air-medium interface, and the ${{x}_{bT}}$ corresponding to the transmitters at different positions can be obtained by solving the Snell's law respectively. ${{k}_{xR}}$ and ${{k}_{z}}$ respectively have a Fourier transform relationship with ${{x}_{R}}$ and ${{z}'}$, and the spatial wave number components ${{k}_{xYT}}$, ${{k}_{xyT}}$, ${{k}_{yR0}}$ and ${{k}_{yR1}}$ respectively satisfy the following equations:
\begin{equation}
	\left\{ \begin{aligned}
		& {{k}_{xYT}}=\sqrt{{{k}^{2}}-{({k}_{z}/2)}^{2}}, \\ 
		& {{k}_{xyT}}=\sqrt{{{k}_{\text{ }\!\!\varepsilon\!\!\text{ }}}^{2}-{({k}_{z}/2)}^{2}}, \\ 
		& {{k}_{yR0}}=\sqrt{{{k}^{2}}-{{{k}_{xR}}^{2}}-{({k}_{z}/2)}^{2}}, \\ 
		& {{k}_{yR1}}=\sqrt{{{k}_{\text{ }\!\!\varepsilon\!\!\text{ }}}^{2}-{{{k}_{xR}}^{2}}-{({k}_{z}/2)}^{2}}. 
	\end{aligned} \right.
	\label{eq: 5.8}
\end{equation}

To facilitate subsequent frequency-domain imaging processing, we first substitute~\eqref{eq: 5.6} and \eqref{eq: 5.7} into~\eqref{eq: 5.1}, and then perform fast Fourier transform (FFT) operations on ${{x}_{R}}$ and ${{z}'}$, to obtain:
\begin{equation}
	\begin{aligned}
		Y(k,{{k}_{xR}},{{k}_{z}},{{x}_{T}}& )=\text{j}{{\eta }_{0}}k\int_{\Omega }{\sigma (\bm{b})\exp (-\text{j}({{k}_{xR}}x+{{k}_{z}}z))}, \\ 
		& \times \exp (-\text{j}{{k}_{xYT}}\sqrt{{{({{x}_{T}}-{{x}_{bT}})}^{2}}+{{\text{Y}}^{2}}}), \\ 
		& \times \exp (-\text{j}{{k}_{xyT}}\sqrt{{({{{x}_{bT}}-x)}^{2}}+{{y}^{2}}}), \\ 
		& \times \exp (\text{j}({{k}_{yR0}}\text{Y}-{{k}_{yR1}}y))\text{d}\bm{b}.  
	\end{aligned}
	\label{eq: 5.9}
\end{equation}

Let
\begin{equation}
	\left\{ \begin{aligned}
		& {{\Phi }_{\text{A}}}=\exp (\text{j}({{k}_{yR1}}y-{{k}_{yR0}}\text{Y})), \\ 
		& {{\Phi }_{\text{B}}}=\exp (\text{j}{{k}_{xYT}}\sqrt{{{({{x}_{T}}-{{x}_{bT}})}^{2}}+{{\text{Y}}^{2}}}), \\ 
		& {{\Phi }_{\text{C}}}=\exp (\text{j}{{k}_{xyT}}\sqrt{{{({{x}_{bT}}-x)}^{2}}+{{y}^{2}}}). \\ 
	\end{aligned} \right.
	\label{eq: 5.10}
\end{equation}

Then~\eqref{eq: 5.9} can be re-expressed as:
\begin{equation}
	\begin{aligned}
		Y(k,{{k}_{xR}},{{k}_{z}},{{x}_{T}}& )=\text{j}{{\eta }_{0}}k\int_{\Omega }{\sigma (\bm{b})\exp (-\text{j}({{k}_{xR}}x+{{k}_{z}}z))} \\ 
		& \text{ }\text{ }\text{ }\times {{\Phi }_{\text{A}}}^{-1} \cdot {{\Phi }_{\text{B}}}^{-1}\cdot {{\Phi }_{\text{C}}}^{-1}
		\text{d}\bm{b}.  
	\end{aligned}
	\label{eq: 5.11}
\end{equation}

According to~\eqref{eq: 5.11}, we first compensate the phase term ${{\Phi }_{\text{A}}}$ for $Y(k,{{k}_{xR}},{{k}_{z}},{{x}_{T}})$, and perform inverse FFT (IFFT) on ${{k}_{xR}}$, we have:
\begin{equation}
	\begin{aligned}
		\tilde{Y}(k,{{x}_{T}},x,y,{{k}_{z}}& )=\text{j}{{\eta }_{0}}k\int_{\Omega }{\sigma (\bm{b})\exp (-\text{j}{{k}_{z}}z)} \\ 
		& \text{ }\text{ }\text{ }\times{{\Phi }_{\text{B}}}^{-1}\cdot {{\Phi }_{\text{C}}}^{-1}
		\text{d}\bm{b}.  
	\end{aligned}
	\label{eq: 5.12}
\end{equation}

Then, the phase terms ${{\Phi }_{\text{B}}}$ and ${{\Phi }_{\text{C}}}$ are compensated for $\tilde{Y}(k,{{x}_{T}},x,y,{{k}_{z}})$, then IFFT is performed on ${{k}_{z}}$ to obtain:
\begin{equation}
\begin{aligned}
	\bar{Y}(k,{{x}_{T}},x,y,z)=\text{j}{{\eta }_{0}}k\int_{\Omega }\sigma (\bm{b}) \text{d}\bm{b}.  
\end{aligned}
	\label{eq: 5.13}
\end{equation}

Finally, by traversing different combinations of spatial wave numbers and transmitting array elements, and coherently accumulating all the target sub-images, the complete MIMO-SAR focused image can be solved, as given by,
\begin{equation}
	\begin{aligned}
		\sigma (\bm{b})=\sum\limits_{{{k}_{m}}}\sum\limits_{{{x}_{Tn}}}{\frac{\text{1}}{\text{j}{{\eta }_{0}}{{k}_{m}}}\bar{Y}(k,{{x}_{T}},x,y,z)},
	\end{aligned}
	\label{eq: 5.14}
\end{equation}
which derives the system model and the problem formulated for our new imaging algorithm demonstrated below.
\section{Proposed Algorithm}\label{sec:design}

In this section, we propose a new algorithm enabling fast and high-quality dielectric target-enhanced imaging. Based on~\eqref{eq: 5.14}, the DT-FDA solution equation applicable to non-uniform MIMO-SAR can be expressed as:
\begin{equation}
	\begin{aligned}
		& \sigma (\bm{b})=\sum\limits_{{{x}_{T}}}{\sum\limits_{k}{\text{IF}{{\text{T}}_{({{k}_{z}})}}}}[\text{IF}{{\text{T}}_{({{k}_{xR}})}}[\text{F}{{\text{T}}_{({{x}_{R}},{z}')}}[ \\ 
		&\text{ } \text{ }\text{ }\text{ }\text{ }\text{ }\text{ }\text{ }\text{ }\text{ }\text{ }\text{ }\text{ }\text{ }\text{ }\text{ }\text{ }\text{ }\text{ }\text{ }\text{ }\text{ }\text{ }\text{ }y(k,{{x}_{T}},{{x}_{R}},{z}')]\cdot {{\Phi }_{\text{A}}}]\cdot {{\Phi }_{\text{B}}}\cdot {{\Phi }_{\text{C}}}].
	\end{aligned}
	\label{eq: 5.15}
\end{equation}

Matrixing the target image that needs to be reconstructed and the four-dimensional spatial target echo data, we can obtain:
\begin{equation}
	\left\{ \begin{aligned}
		& \textbf{\text{H}}=\sigma (\bm{b}), \\ 
		& \textbf{\text{Y}}=y(k,{{x}_{T}},{{x}_{R}},{z}'). \\ 
	\end{aligned} \right.
	\label{eq: 5.16}
\end{equation}

The imaging process corresponding to~\eqref{eq: 5.15} can be represented in the following matrix operation form:
\begin{equation}
	\begin{aligned}
		\textbf{\text{H}}=\Psi{{}^{\dagger }}\{\textbf{\text{Y}}\}=\sum\limits_{{{x}_{T}}}{\sum\limits_{k}{(\bar{\textbf{\text{F}}}(\textbf{\text{F}}\textbf{\text{Y}}{{\textbf{\text{F}}}^{T}}\circ {{\Phi }_{\text{A}}})\circ {{\Phi }_{\text{B}}}\circ {{\Phi }_{\text{C}}}){{\textbf{\text{F}}}^{H}}}},
	\end{aligned}
	\label{eq: 5.17}
\end{equation}
where $\Psi{{}^{\dagger }}$ denotes the established inverse sensing operator, and $\{\cdot\}$ constitutes the input entity of the operator function. $\textbf{\text{F}}$  represents the Fourier transform matrix, while $\bar{\textbf{\text{F}}}$, ${{\textbf{\text{F}}}^{T}}$, and ${{\textbf{\text{F}}}^{H}}$ represent the conjugate, transpose, and conjugate transpose of $\textbf{\text{F}}$, respectively. $\circ$ represents the Hadamard product, which is the matrix point multiplication operation symbol. Equation~\eqref{eq: 5.17} shows that in the non-uniform MIMO-SAR system, with the echo matrix $\textbf{\text{Y}}$ utilized as the input of the inverse sensing operator function, the internal structure of the half-space dielectric target will be quickly reconstructed.

Then, by inversely deducing the target echo matrix $\textbf{\text{Y}}$ according to \eqref{eq: 5.17}, the forward sensing operator $\Psi$ can be easily constructed, that is:
\begin{equation}
	\begin{aligned}
		\textbf{\text{Y}}=\Psi\{\textbf{\text{H}}\}=\sum\limits_{y}{\bar{\textbf{\text{F}}}(\textbf{\text{F}}(\textbf{\text{H}}{{\textbf{\text{F}}}^{T}}\circ {{{\bar{\Phi }}}_{\text{B}}}\circ {{\Phi }_{\text{C}}})\circ {{{\bar{\Phi }}}_{\text{A}}}){{\textbf{\text{F}}}^{H}}},
	\end{aligned}
	\label{eq: 5.18}
\end{equation}
where ${{\bar{\Phi }}_{\text{A}}}$, ${{\bar{\Phi }}_{\text{B}}}$ and ${{\bar{\Phi }}_{\text{C}}}$ are the conjugates of ${{\Phi }_{\text{A}}}$, ${{\Phi }_{\text{B}}}$ and ${{\Phi }_{\text{C}}}$, respectively. 

In the following, the detailed derivation of the fast iterative image-solving process based on the ADMM framework will be presented according to the forward sensing operator constructed above.

Let $\text{vec}(\cdot)$ denote the vectorization operator, and through the vectorization representation of the echo matrix, the non-uniform MIMO-SAR echo model can be presented in the subsequent linear form:
\begin{equation}
	\begin{aligned}
		\textbf{\text{y}}=\textbf{\text{A}}\textbf{\text{h}},
	\end{aligned}
	\label{eq: 5.19}
\end{equation}
where $\textbf{\text{y}}=\text{vec}(\textbf{\text{Y}})$, and $\textbf{\text{h}}=\text{vec}(\textbf{\text{H}})$. $\textbf{\text{A}}$ represents the extensive sensing matrix, and its matrix elements are derived from ${{A}_{TR}}(k,\bm{b},{\bm{b}_{T}},{\bm{b}_{R}})$.

To increase the sparsity of the target image and achieve fast convergence simultaneously, the imaging solution can often be formulated as the subsequent linear least-squares problem (unconstrained CS optimization problem) with an ${{\ell }_{1}}$ regularization term:
\begin{equation}
	\begin{aligned}
		\hat{\textbf{\text{h}}}=\text{arg} \underset{\textbf{\text{h}}}{\mathop{\text{ min}}}\,\text{ }\frac{1}{2}\left\| \textbf{\text{y}}-\textbf{\text{A}}\textbf{\text{h}} \right\|_{2}^{2}+\lambda {{\left\| \textbf{\text{h}} \right\|}_{1}},
	\end{aligned}
	\label{eq: 4.6}
\end{equation}
where ${{\left\| \cdot  \right\|}_{2}}$ represents the ${{\ell }_{2}}$ (Euclidean) norm, which can be utilized to evaluate the precision and firmness of the optimization outcome. ${{\left\| \cdot  \right\|}_{1}}$ stands for the ${{\ell }_{1}}$ norm, which indicates the reconstructed image's sparsity. $\lambda$ serves as a regularization parameter to weigh the two separate penalty terms within the objective function.

In light of the ADMM theory\cite{33}, by introducing an extra auxiliary variable $\textbf{\text{r}}$, \eqref{eq: 4.6} can be further developed into the subsequent constrained optimization issue:
\begin{equation}
	\begin{aligned}
		& \hat{\textbf{\text{h}}}=\text{arg} \underset{\textbf{\text{h}}}{\mathop{\text{ min}}}\,\text{ }\frac{1}{2}\left\| \textbf{\text{y}}-\textbf{\text{A}}\textbf{\text{h}} \right\|_{2}^{2}+\lambda {{\left\| \textbf{\text{r}} \right\|}_{1}}, \\ 
		& \text{s}\text{.t}\text{.  }\textbf{\text{h}}=\textbf{\text{r}}.
	\end{aligned}
	\label{eq: 4.7}
\end{equation}

According to~\eqref{eq: 4.7}, the following augmented Lagrangian function can be established, i.e.,
\begin{equation}
	\begin{aligned}
		& {{\text{L}}_{\rho }}\left( \textbf{\text{h}},\textbf{\text{r}},\textbf{\text{m}} \right)=\frac{1}{2}\left\| \textbf{\text{y}}-\textbf{\text{A}}\textbf{\text{h}} \right\|_{2}^{2}+\lambda {{\left\| \textbf{\text{r}} \right\|}_{1}} \\
		& \text{ }\text{ }\text{ }\text{ }\text{ }\text{ }\text{ }\text{ }\text{ }\text{ }\text{ }\text{ }\text{ }\text{ }\text{ }\text{ }+{{\textbf{\text{m}}}^{H}}\left( \textbf{\text{h}}-\textbf{\text{r}} \right)+\frac{\rho }{2}\left\| \textbf{\text{h}}-\textbf{\text{r}} \right\|_{2}^{2}, 
	\end{aligned}
	\label{eq: 4.8}
\end{equation}
where $\textbf{\text{m}}$ and $\rho$ represent the Lagrange multiplier and penalty parameter, and ${{\textbf{\text{m}}}^{H}}$ is the conjugate transpose of $\textbf{\text{m}}$.

By using the matrix operator $\text{unvec}(\cdot )$, the vectorized function given in~\eqref{eq: 4.8} can be expressed in matrix form:
\begin{equation}
	\begin{aligned}
		{{\text{L}}_{\rho }}\left( \textbf{\text{H}},\textbf{\text{R}},\textbf{\text{M}} \right)&=\frac{1}{2}\left\| \textbf{\text{Y}}-\text{unvec(}\textbf{\text{A}}\text{vec}(\textbf{\text{H}})\text{)} \right\|_{2}^{2}+\lambda {{\left\| \textbf{\text{R}} \right\|}_{1}} \\ 
		& \text{ }\text{ }+\text{Tr}\left\{ {{\textbf{\text{M}}}^{H}}\left( \textbf{\text{H}}-\textbf{\text{R}} \right) \right\}+\frac{\rho }{2}\left\| \textbf{\text{H}}-\textbf{\text{R}} \right\|_{2}^{2}, 
	\end{aligned}
	\label{eq: 4.9}
\end{equation}
where $\textbf{\text{R}}$ and $\textbf{\text{M}}$ respectively stand for the matrixized auxiliary variable and dual variable. $\text{Tr}\left\{ \cdot  \right\}$ is utilized to solve the matrix's trace.

By substituting the forward sensing operator $\Psi$ in~\eqref{eq: 5.18} for the extensive sensing matrix $\textbf{\text{A}}$ in~\eqref{eq: 4.9}, as given by:
\begin{equation}
	\begin{aligned}
		{{\text{L}}_{\rho }}\left( \textbf{\text{H}},\textbf{\text{R}},\textbf{\text{M}} \right)&=\frac{1}{2}\left\| \textbf{\text{Y}}-\Psi\{\textbf{\text{H}}\} \right\|_{2}^{2}+\lambda {{\left\| \textbf{\text{R}} \right\|}_{1}} \\ 
		& \text{ }\text{ }+\text{Tr}\left\{ {{\textbf{\text{M}}}^{H}}\left( \textbf{\text{H}}-\textbf{\text{R}} \right) \right\}+\frac{\rho }{2}\left\| \textbf{\text{H}}-\textbf{\text{R}} \right\|_{2}^{2}. 
	\end{aligned}
	\label{eq: 4.10}
\end{equation}

Since the non-uniform MIMO-SAR DT-FDA only uses the FFT/IFFT, matrix product, and cumulative steps in the execution process, it is considered that the sensing operator functions $\Psi$ and $\Psi{{}^{\dagger }}$ satisfy the linearity and orthogonality properties \cite{27,29}.

Taking the first-order partial derivative of ${{\text{L}}_{\rho }}\left( \textbf{\text{H}},\textbf{\text{R}},\textbf{\text{M}} \right)$ in relation to $\textbf{\text{H}}$, we can obtain:
\begin{equation}
	\begin{aligned}
		\frac{\partial {{\text{L}}_{\rho }}\left( \textbf{\text{H}},\textbf{\text{R}},\textbf{\text{M}} \right)}{\partial \textbf{\text{H}}}=-\Psi{{}^{\dagger }}\{\textbf{\text{Y}}-\Psi\{\textbf{\text{H}}\}\}+\textbf{\text{M}}+\rho (\textbf{\text{H}}-\textbf{\text{R}}).
	\end{aligned}
	\label{eq: 4.11}
\end{equation}

Let~\eqref{eq: 4.11} be zero to obtain the analytical expression of $\textbf{\text{H}}$, that is:
\begin{equation}
	\begin{aligned}
		\textbf{\text{H}}=\frac{\Psi{{}^{\dagger }}\{\textbf{\text{Y}}\}+\rho \textbf{\text{R}}-\textbf{\text{M}}}{1+\rho }.
	\end{aligned}
	\label{eq: 4.12}
\end{equation}

Similarly, using the quadratic convex optimization theory, and then taking the first-order partial derivative of ${{\text{L}}_{\rho }}\left( \textbf{\text{H}},\textbf{\text{R}},\textbf{\text{M}} \right)$ regarding $\textbf{\text{R}}$, the analytical formulation of $\textbf{\text{R}}$ will be obtained, as given by,
\begin{equation}
	\begin{aligned}
		\textbf{\text{R}}=\mathcal{T}(\textbf{\text{H}}+\frac{\textbf{\text{M}}}{\rho };\frac{\lambda }{\rho }).
	\end{aligned}
	\label{eq: 4.13}
\end{equation}

In~\eqref{eq: 4.13}, $\mathcal{T}(\textbf{\text{U}};v)$ symbolizes the complex soft-thresholding function, and its separate element is defined as:
\begin{equation}
	\begin{aligned}
		{{[\mathcal{T}(\textbf{\text{U}};v)]}_{ij}}=\frac{{{\textbf{\text{U}}}_{ij}}}{\left| {{\textbf{\text{U}}}_{ij}} \right|}\cdot \max (\left| {{\textbf{\text{U}}}_{ij}} \right|-v;\text{ 0}),
	\end{aligned}
	\label{eq: 4.14}
\end{equation}
where ${{\textbf{\text{U}}}_{ij}}$ represents the value of $\textbf{\text{U}}$ at the position of $i$-th row and $j$-th column, and $\left| \cdot  \right|$ is used for calculating the element's absolute value.

Since the solution process for $\textbf{\text{R}}$ is the same as that given in references \cite{29,30,34}, the specific derivation procedure will not be reiterated here.

Let $t$ represent the number of iteration steps, then the iterative format of the optimized ADMM for reconstructing the target image is as follows:
\begin{equation}
	\left\{ \begin{aligned}
		& {{\textbf{\text{H}}}^{(t)}}=\frac{\Psi{{}^{\dagger }}\{\textbf{\text{Y}}\}+\rho {{\textbf{\text{R}}}^{(t-1)}}-{{\textbf{\text{M}}}^{(t-1)}}}{1+\rho }, \\ 
		& {{\textbf{\text{R}}}^{(t)}}=\mathcal{T}({{\textbf{\text{H}}}^{(t)}}+\frac{{{\textbf{\text{M}}}^{(t-1)}}}{\rho };\frac{\lambda }{\rho }), \\ 
		& {{\textbf{\text{M}}}^{(t)}}={{\textbf{\text{M}}}^{(t-1)}}+\rho ({{\textbf{\text{H}}}^{(t)}}-{{\textbf{\text{R}}}^{(t)}}). 
	\end{aligned} \right.
	\label{eq: 4.15}	
\end{equation}

The validity of the proposed ADMM iterative solution equation relies on the choice of the regularization parameter $\lambda$, which requires corresponding adjustment in accordance with diverse imaging scenarios, target, array, and system parameter settings. 
Our previous work~\cite{35} can be applied by adopting the optimal regularization parameter search method for parameter optimization, and simultaneously dynamically regulate the quantity of circular iterations needed for ADMM convergence to further improve the image reconstruction quality of the internal structure or defects of the dielectric material while achieving high computational efficiency.
\section{Analysis of Measured Data Imaging Results}\label{sec:evaluation}

This section utilizes two distinct sets of non-uniform MIMO-SAR measured data to evaluate the robustness of the proposed enhanced imaging algorithm in real-world applications. The evaluation focuses on the performance of the algorithm in terms of imaging accuracy and computational efficiency, as well as its adaptability to different non-uniform MIMO array layouts and diverse imaging targets.

\subsection{PTFE Block with Four Cavities Inside}
\label{sec:exp_1}

\begin{figure}[t!]
	\centering
	\subfloat[]{
		\includegraphics[width=0.26\linewidth]{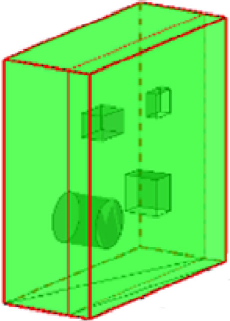}
	}\text{ }\text{ }\text{ }\text{ }\text{ }\text{ }\text{ }\text{ }\text{ }\text{ }
	\subfloat[]{
		\includegraphics[width=0.295\linewidth]{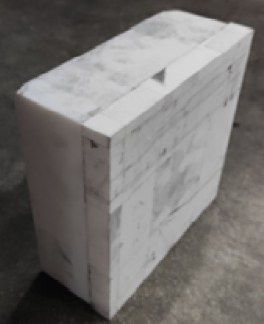}
	}
	\caption{PTFE block with four internal cavities. (a) Conceptual schematic diagram. (b) Physical picture. }
	\label{fig: 5.7}
\end{figure}

\begin{figure}[t!]
	\centering
	\includegraphics[width=0.75\linewidth]{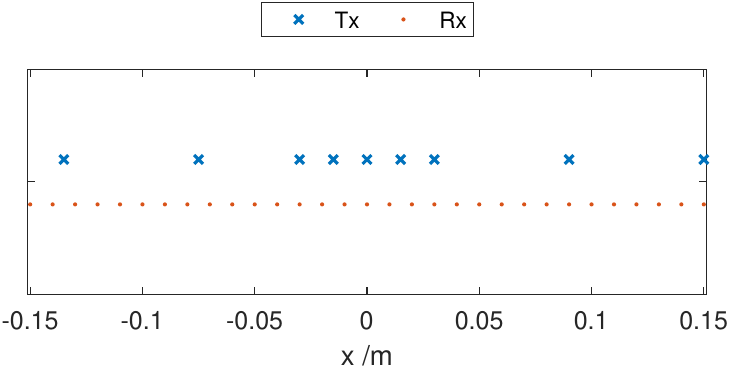}
	\caption{The non-uniform MIMO array used in the experiment of Section~\ref{sec:exp_1}}
	\label{fig: 5.3}
\end{figure}

For simulating the structural defects within the dielectric target, in this experiment, a PTFE block containing four cavities is utilized as the imaging target. As shown in Fig. \ref{fig: 5.7}, the target's size is 0.2 m $\times$ 0.08 m $\times$ 0.2 m, and it contains one cylindrical cavity and three cube cavities of different sizes. The non-uniform MIMO array configuration employed in the experiment is depicted in Fig. \ref{fig: 5.3}, which consists of 9 non-uniformly arranged transmitting elements and 31 evenly distributed receiving elements, and the array has a total length of 0.3 m. The mechanical scanning length and scanning pitch in the height direction are 0.228 m and 0.003 m, respectively, and the entire MIMO-SAR array aperture is placed on the 2D plane where $\text{Y}=-\text{0.3 m}$. Supposing that the operating frequency of the system ranges from 31.5$\sim$43.5 GHz with 51 sampling frequency points, by scanning point by point with a 2D mechanical scanner and a vector network analyzer (VNA), the complete MIMO-SAR spatial target echo signal of 9 $\times$ 31 $\times$ 77 $\times$ 51 can be obtained.

\begin{figure}[t!]
	\centering
	\subfloat[]{
		\includegraphics[width=0.47\linewidth]{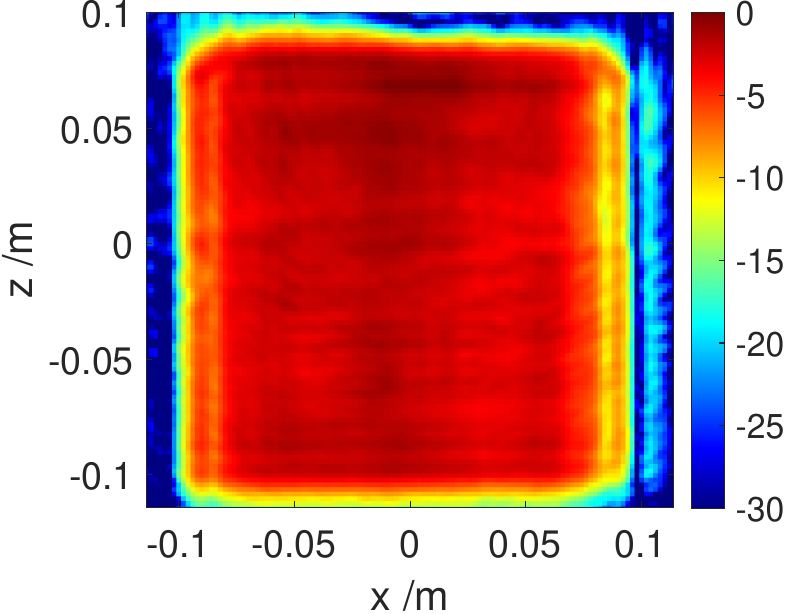}\text{ }\text{ }\includegraphics[width=0.47\linewidth]{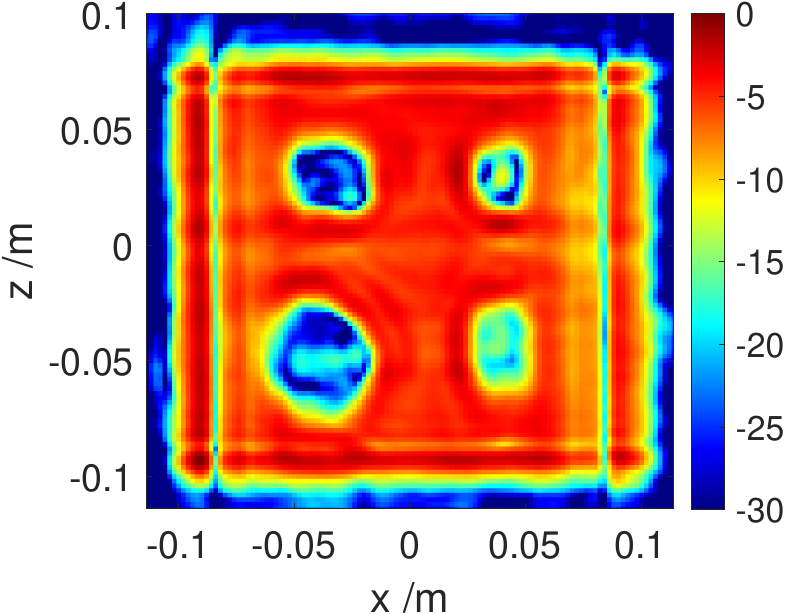}
	}
	\\
	\subfloat[]{
		\includegraphics[width=0.47\linewidth]{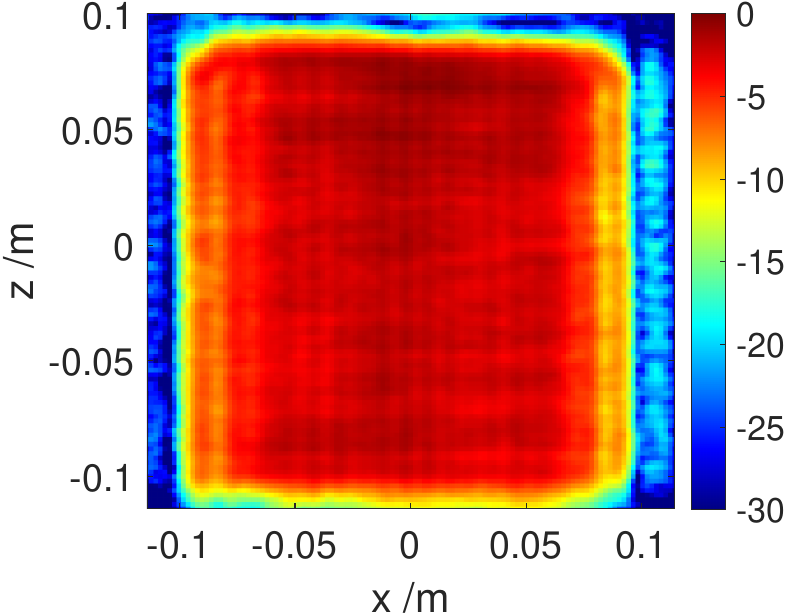}\text{ }\text{ }\includegraphics[width=0.47\linewidth]{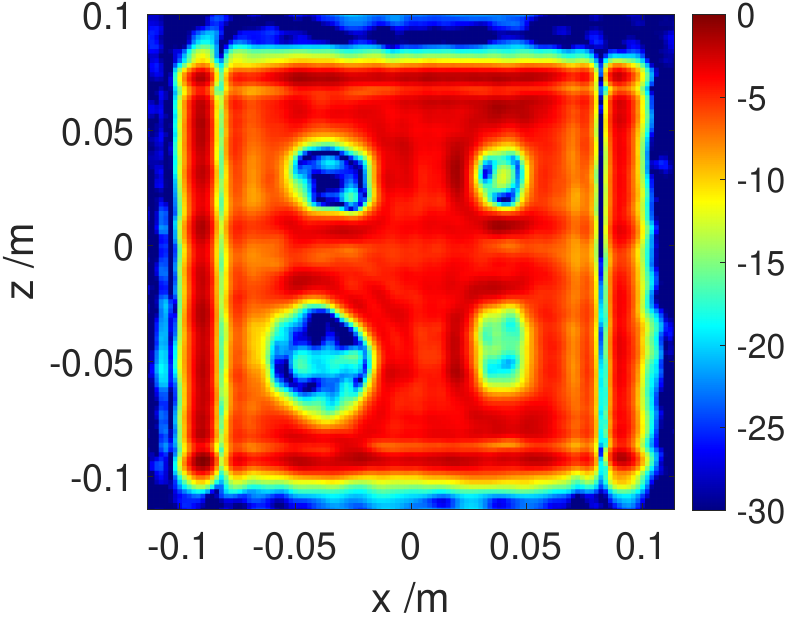}
	}
	\\
	\subfloat[]{
		\includegraphics[width=0.47\linewidth]{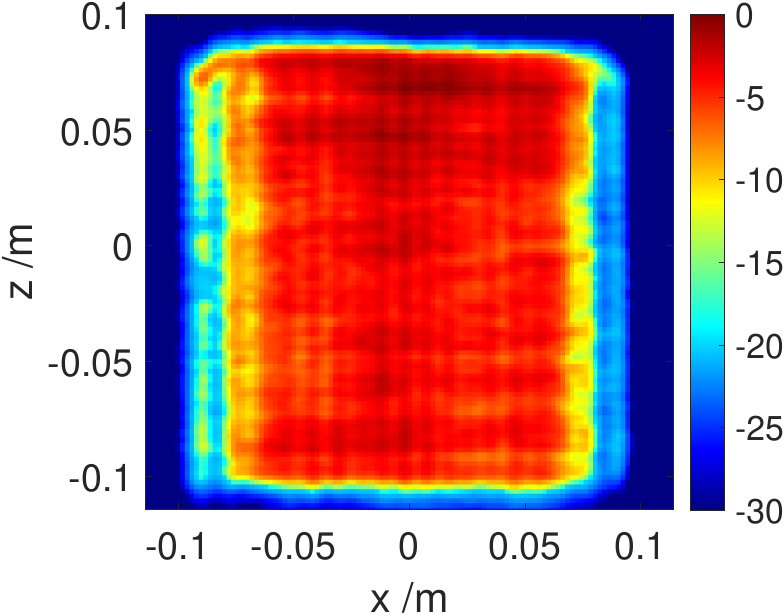}\text{ }\text{ }\includegraphics[width=0.47\linewidth]{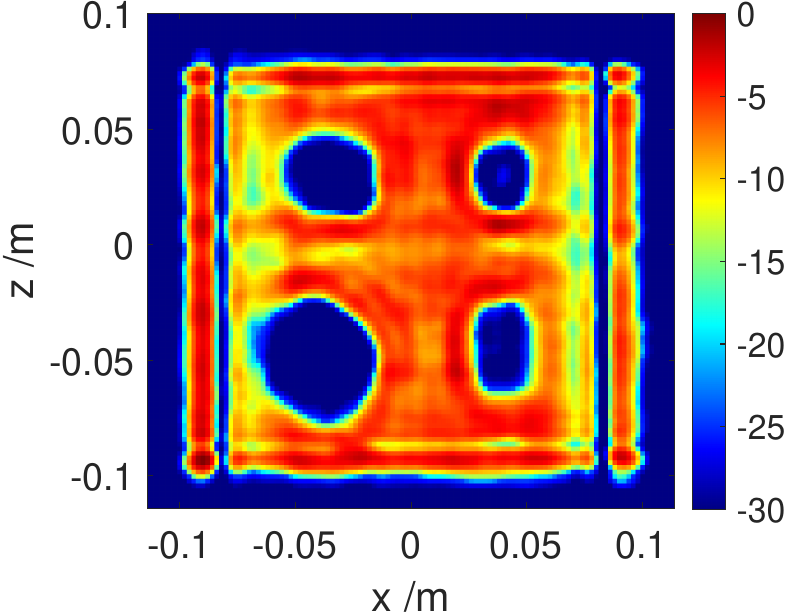}
	}
	\caption{Sectional imaging results of PTFE block at different distances through different algorithms. (a) IBP algorithm. (b) DT-FDA. (c) Enhanced algorithm.}
	\label{fig: 5.10}
\end{figure}

Fig. \ref{fig: 5.10} depicts the 2D sectional imaging results at different distances obtained by processing the measured data through three algorithms, where the planes $y=0 \text{ cm}$ and $y=8 \text{ cm}$ in line with the front surface (i.e., air medium interface) and the rear surface of the PTFE block, respectively. Since the imaging process at $y=0 \text{ cm}$ is not influenced by the internal structure of the medium, it can be regarded as imaging the front surface of the PTFE block in free space. Apparently, the three algorithms all possess the capability to reconstruct the surface of the medium structure. When $y=8 \text{ cm}$, defects corresponding to the cavity positions appear in the rear surface of the PTFE block for all three algorithms. The reason for this phenomenon is that the numerous scattering and diffraction of electromagnetic waves at the cavity position will lead to the attenuation of forward-propagating electromagnetic waves, thereby reducing the scattering intensity of the rear surface of the PTFE block placed directly behind. However, it can be observed from the results that, compared to the IBP algorithm and DT-FDA, the enhanced algorithm achieves superior focusing and sidelobe suppression performance, while effectively suppressing the non-ideal scattering effects existing in the imaging region, significantly improving the imaging quality. This also provides favorable conditions for the subsequent research on defect detection and identification of the dielectric target's internal structure.

\begin{table}[t!]
	\setlength{\tabcolsep}{15pt}
	\caption{The IE Values of Imaging Results Corresponding to Different Algorithms in the experiment of Section~\ref{sec:exp_1}}
	\centering
	\label{table: 5.5}
	\begin{tabular}{@{}cccc@{}}
		\toprule
		& $y=0\text{ cm}$ & $y=8\text{ cm}$ \\ \midrule
		IBP algorithm & 9.095 & 8.954 \\
		DT-FDA & 9.103 & 8.946 \\ 
		Enhanced algorithm & \textbf{8.852} & \textbf{8.664} \\ 
		\bottomrule
	\end{tabular}
\end{table}

To quantitatively evaluate the focusing performance of reconstructed target images using different algorithms, Table \ref{table: 5.5} presents the image entropy (IE) values of the experimental imaging results corresponding to the three algorithms. For the 2D sections at different distances within the PTFE block, the IE of the enhanced algorithm is significantly lower than that of the other algorithms, further verifying the good focusing performance of the algorithm and its suppression ability on the non-ideal scattering effects caused by the dielectric material. Regarding computing efficiency, the imaging time required to execute the IBP algorithm is 23295.3 s, while the imaging time of the enhanced algorithm is only 15.29 s, which is close to the imaging time of 14.51 s corresponding to DT-FDA, achieving a similar high imaging computational efficiency.

\vspace{2pt}
\begin{center}
\fbox{%
    \begin{minipage}{0.96\linewidth}

    \textbf{Takeaway--superior performance and faster.} The enhanced algorithm demonstrates superior focusing and sidelobe suppression, effectively suppresses non-ideal scattering effects, and achieves significantly lower image entropy compared to IBP and DT-FDA. It also offers drastically improved computational efficiency, with imaging time reduced from 23295.3s (IBP) to just 15.29s, comparable to DT-FDA's 14.51s.
    \end{minipage}
}
\end{center}

\subsection{PTFE Block Embedded with Metal Lemon Slice}
\label{sec:exp_2}

\begin{figure}[t!]
	\centering
	\subfloat[]{
		\includegraphics[width=0.22\linewidth]{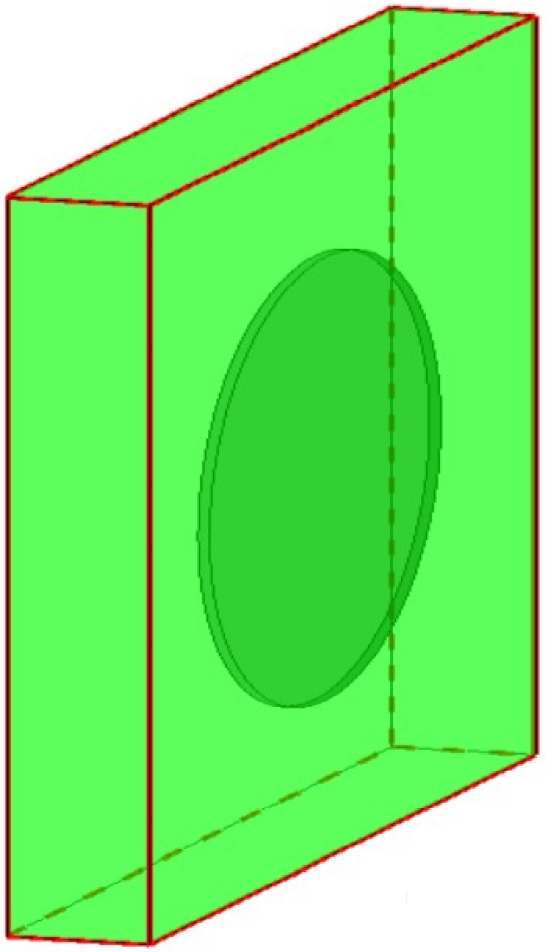}
	}\text{ }\text{ }\text{ }\text{ }\text{ }\text{ }\text{ }\text{ }\text{ }
	\subfloat[]{
		\includegraphics[width=0.3\linewidth]{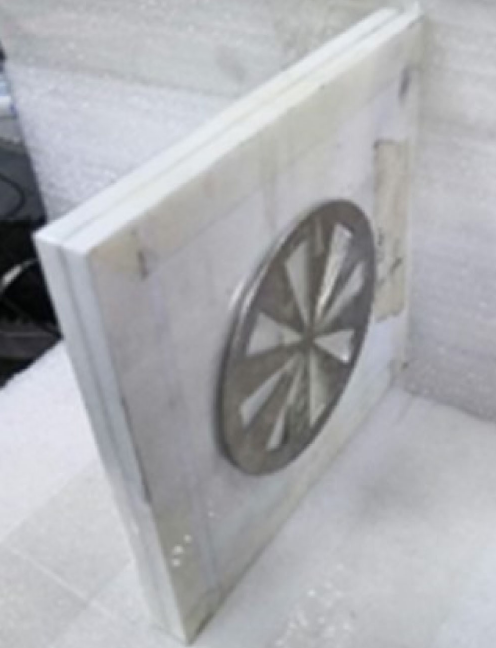}
	}
	\caption{PTFE block with embedded metal lemon slice. (a) Conceptual schematic diagram. (b) Physical picture.}
	\label{fig: 5.11}
\end{figure}

\begin{figure}[t!]
	\centering
	\includegraphics[width=0.75\linewidth]{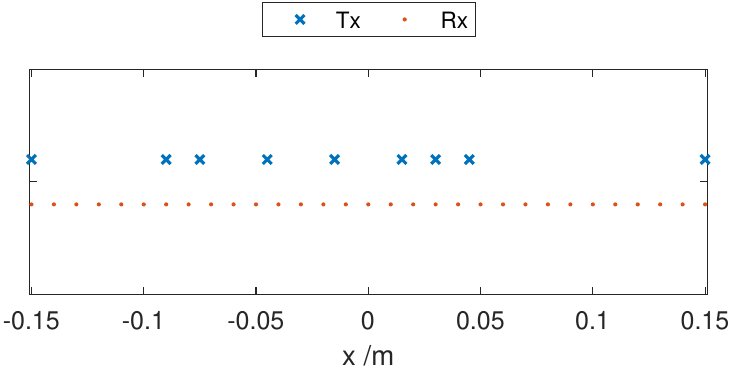}
	\caption{The non-uniform MIMO array used in the experiment of Section~\ref{sec:exp_2}}
	\label{fig: 5.12}
\end{figure}

In this experiment, PTFE block with embedded metal lemon slice presented in Fig. \ref{fig: 5.11} is employed as the imaging object. At the same time, the position arrangement of the transmitting elements in the MIMO array is partially adjusted to verify the stability of enhanced algorithm for different array configurations and imaging targets. The adjusted non-uniform MIMO array is depicted in Fig. \ref{fig: 5.12}, which also contains 9 non-uniformly arranged transmitting elements and 31 evenly distributed receiving elements, with a total length of 0.3 m. In addition, the mechanical scanning length is set to 0.36 m. The remaining experimental parameters, such as the signal operating frequency, the number of sweep points, and the SAR scanning interval, are all exactly the same as that of the evaluation results in Section~\ref{sec:exp_1}.

\begin{figure}[t!]
	\centering
	\subfloat[]{
		\includegraphics[width=0.52\linewidth]{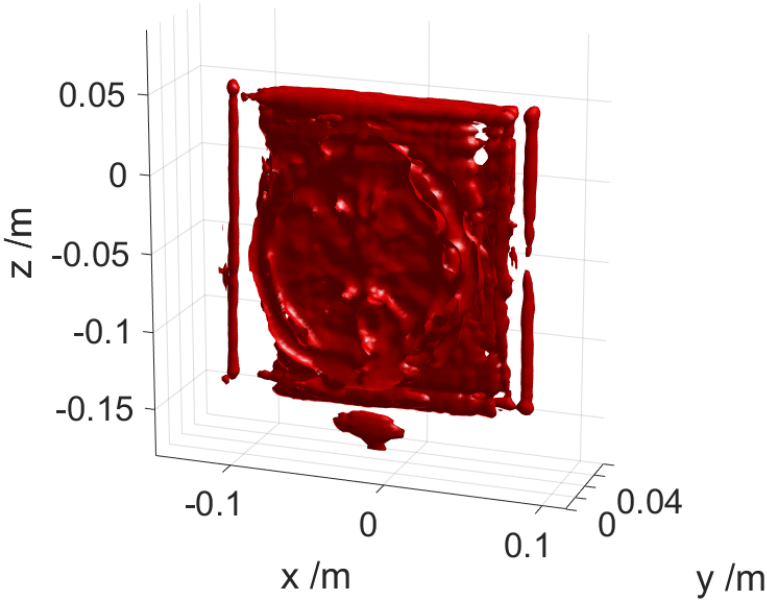}\includegraphics[width=0.46\linewidth]{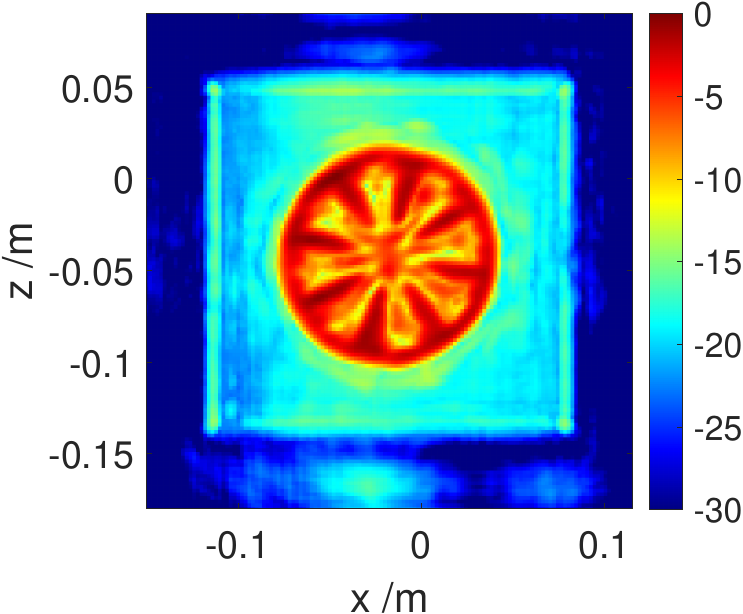}
	}
	\\
	\subfloat[]{
		\includegraphics[width=0.52\linewidth]{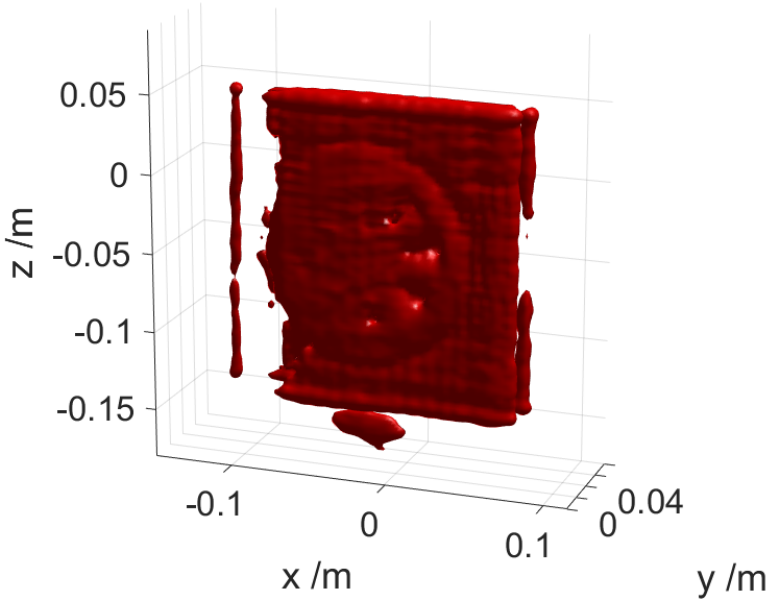}\includegraphics[width=0.46\linewidth]{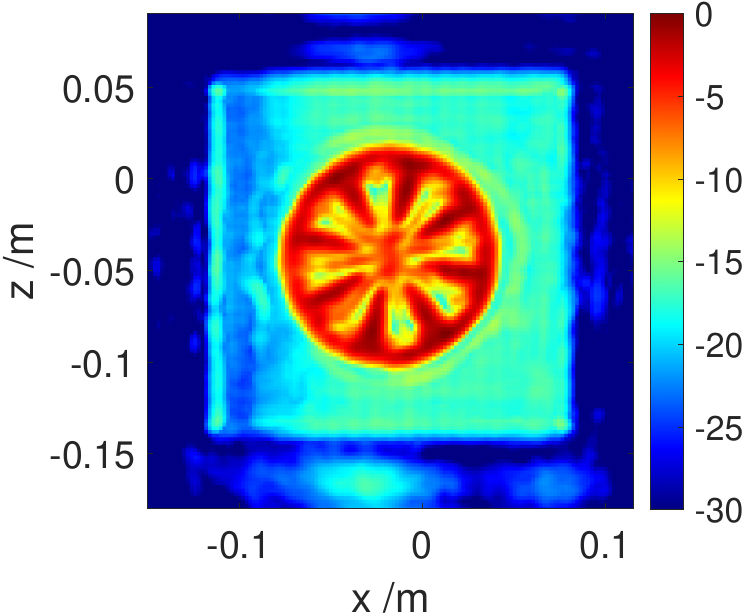}
	}
	\\
	\subfloat[]{
		\includegraphics[width=0.52\linewidth]{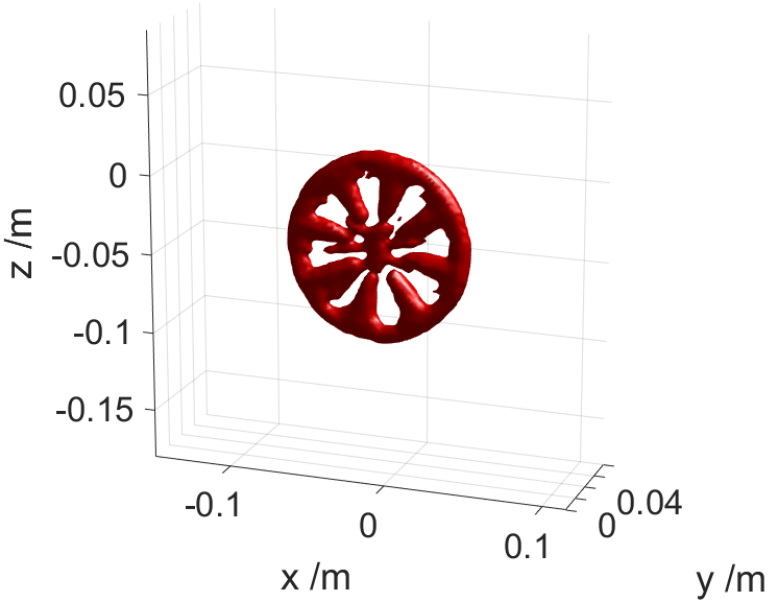}\includegraphics[width=0.46\linewidth]{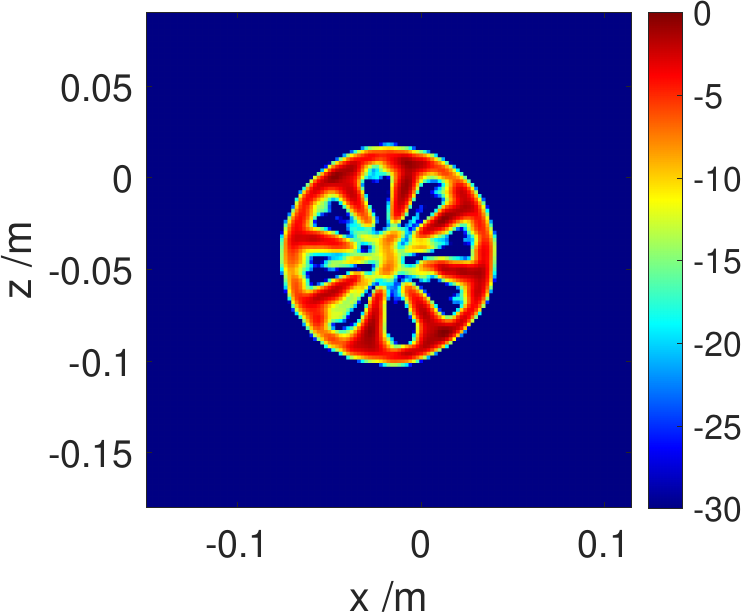}
	}
	\caption{Imaging results of the metal lemon slice inside PTFE block using different algorithms. (a) IBP algorithm. (b) DT-FDA. (c) Enhanced algorithm.}
	\label{fig: 5.13}
\end{figure}

Fig. \ref{fig: 5.13} depicts the 3D imaging results of the measured echo data using the IBP algorithm, DT-FDA, and the enhanced algorithm, as well as their maximum projections in the $y$ direction. The voxel intensity of the curved surface wrapped area in the 3D view is greater than $-$20 dB, and the displayed dynamic range of the matching maximum projection result is 30 dB. It is obvious from the results that due to the influence of the non-ideal scattering effect caused by the dielectric material, the contour and shape details of the metal lemon slice in the 3D views corresponding to the IBP algorithm and DT-FDA can no longer be identified, while the enhanced algorithm can effectively suppress the weak scattering components and achieve high-precision 3D reconstruction of the strong scattering targets inside the dielectric. In addition, it can also be found from the maximum projection results that the enhanced algorithm exhibits higher focusing and sidelobe suppression abilities compared to the other two algorithms, which fully demonstrates the potential application value of the enhanced algorithm in concealed object detection.

\begin{table}[t!]
	\setlength{\tabcolsep}{10pt}
	\caption{The Comparison of IE and Imaging Time of Different Algorithms in the experiment of Section~\ref{sec:exp_2}}
	\centering
	\label{table: 5.6}
	\begin{tabular}{@{}cccc@{}}
		\toprule
		& IE & Imaging Time (s) \\ \midrule
		IBP algorithm & 11.248 & 40328.60 \\ 
		DT-FDA & 11.146 & 14.86 \\
		Enhanced algorithm & \textbf{8.864} & \textbf{16.60} \\ 
		\bottomrule
	\end{tabular}
\end{table}

Table \ref{table: 5.6} presents the IE corresponding to different algorithms. As expected, the IE of enhanced algorithm is only 8.864, considerably less than that of the IBP algorithm and DT-FDA. The imaging time of three algorithms is also listed in Table \ref{table: 5.6}. Due to the similar parameter settings, the imaging time laws of three algorithms are consistent with the that of experiments in Section~\ref{sec:exp_1}. The enhanced algorithm still maintains the smallest IE and high computational efficiency, which further confirms its application potential in practical scenarios. If GPU parallel acceleration technology is adopted, the enhanced algorithm can achieve similar imaging efficiency to DT-FDA, thereby achieving high-precision real-time enhanced imaging.

\vspace{2pt}
\begin{center}
\fbox{%
    \begin{minipage}{0.96\linewidth}

    \textbf{Takeaway--more accurate and efficient.} The enhanced algorithm outperforms IBP and DT-FDA in 3D imaging of a PTFE block with embedded metal, showing better shape reconstruction, higher focusing ability, and stronger sidelobe suppression. It achieves the lowest Image Entropy (8.864) and maintains high computational efficiency, with potential for real-time enhanced imaging using GPU acceleration.
    \end{minipage}
}
\end{center}
\section{Related Work}\label{sec:related}

\subsection{Temporal-domain algorithms}
As a conventional temporal domain imaging algorithm, the IBP algorithm \cite{19} boasts intuitive physical meaning, simple algorithm processing, and high imaging accuracy, and is suitable for nearly all MIMO antenna array arrangements, making it an ``all-purpose'' dielectric target imaging algorithm. However, in the MMW frequency band, to achieve high-resolution imaging, the amount of mesh points divided within the same imaging region far exceeds that required for microwave imaging. Coupled with the high-dimensional target echo data recorded by the MIMO-SAR system and the numerous refractive point position calculation steps, the amount of computation required to execute this algorithm is substantial, resulting in very low imaging efficiency.

\subsection{Frequency-domain algorithms}
\noindent\textbf{Uniform MIMO array.}
Deng et al. introduced a fast and fully focused MIMO-SAR dielectric target imaging algorithm that ensures accurate reconstruction of the internal structure of dielectric targets while substantially improving imaging efficiency \cite{20}. Their approach utilizes frequency-domain processing on half-space dyadic Green's function, enabling high-quality and swift imaging of dielectric targets' internal structures through the application of FFT and phase compensation techniques. However, this algorithm's effectiveness is contingent upon the use of a strictly uniform transmitting and receiving array within the MIMO system, limiting its applicability to non-uniform MIMO array configurations.

\smallskip
\noindent\textbf{Non-uniform MIMO array.}
Different from the uniform MIMO array system, some optimized non-uniform MIMO array topologies, such as the real-aperture linear MIMO array with the \textit{uniform receiving and non-uniform transmitting} or \textit{uniform transmitting and non-uniform receiving} have better grating sidelobe suppression capabilities \cite{21,22,23,24}. Therefore, investigating an efficient algorithm for non-uniform MIMO-SAR dielectric targets imaging has certain theoretical significance and practical application value.

\smallskip
\noindent\textbf{DT-FDA.}
Chen et al. \cite{25} developed a method for imaging half-space layered dielectric targets in non-uniform MIMO-SAR systems. They analyzed spatial wave numbers in different media, used partial FFT and spherical wave decomposition for accurate frequency-domain echo expression, and quickly reconstructed the target image through IFFT, phase compensation, and wave number integration. However, their algorithm (DT-FDA) suffers from clutter and sidelobe effects in high dynamic ranges due to aperture and bandwidth limitations. Additionally, non-ideal scattering from dielectric materials impacts subsequent target recognition and image interpretation.

\subsection{CS-based algorithms.}
Although the imaging approach based on CS methods can significantly improve the quality of target imaging \cite{26,27}, most of the existing CS-based basis pursuit algorithms such as ADMM require storing and handling massive measurement matrices in the iterative solving procedure, which poses extremely high demands on the performance of computing device performance \cite{28,33}.

\subsection{What Sets Our Solution Apart}
Existing fast imaging techniques encounter image quality problems at high dynamic range, and most of the CS methods need to handle massive matrices, which requires high computing resources. Consequently, our new algorithm integrates DT-FDA into an ADMM-based iterative solution process, avoiding large-scale matrix operations and reducing computing resource requirements while maintaining high efficiency. Moreover, the algorithm's good sparse performance leads to an effective improvement in image quality.

\section{Conclusion}\label{sec:conclusion}
This paper presented a fast and high-quality dielectric target-enhanced imaging algorithm. This algorithm introduces the existing DT-FDA into the iterative solution process based on ADMM, effectively suppressing the non-ideal scattering effect caused by the dielectric material, and realizing enhanced imaging of the internal structure or defect of the layered dielectric under the foundation of guaranteeing high computational efficiency. 
The proposed algorithm demonstrates superior performance across scenarios, achieving the lowest image entropy (8.864) compared to IBP and DT-FDA. It significantly improves efficiency, reducing imaging time from 23295.3 s (IBP) to 15.29 s, comparable to DT-FDA's 14.51 s. These improvements highlight its potential for advanced defect detection and target recognition in dielectric structures.

\bibliographystyle{IEEEtran}
\bibliography{Ref}
\end{document}